\documentclass{iau} 

\usepackage{graphicx}
\usepackage{xcolor}
\usepackage[hyphens]{url}
\usepackage{hyperref}
\hypersetup{
     colorlinks  = true,
     linkcolor   = blue, 
     anchorcolor = BrickRed,
     citecolor   = blue,
     filecolor   = blue,
     menucolor   = blue,
     runcolor    = cyan,
     urlcolor    = blue
}
\usepackage[authoryear]{natbib}

\pubyear{2019}
\volume{355} 
\jname{The Realm of the Low-Surface-Brightness Universe}
\editors{D. Valls-Gabaud, I. Trujillo \& S. Okamoto, eds.}

\def\etal{\textit{et al.}}

\title[Rings and spiral arms] 
{Rings and spiral arms: \\ are they coupled with bars?}

\author[Sim\'on D\'iaz-Garc\'ia et al.]   
{Sim\'on D\'iaz-Garc\'ia$^{1,2}$, 
Johan H. Knapen$^{1,2}$, 
Heikki Salo$^{3}$, 
Mart\'in Herrera-Endoqui$^{4}$,
\and Sergio D\'iaz-Su\'arez$^{1,2}$
}

\affiliation{
$^1$Instituto de Astrof\'isica de Canarias, 
E-38205, La Laguna, Tenerife, Spain \\ email: {\tt simondiazgar@gmail.com} \\[\affilskip]
$^2$Departamento de Astrof\'isica, Universidad de La Laguna, E-38205, La Laguna, Tenerife, Spain\\[\affilskip]
$^3$Astronomy Research Unit, University of Oulu, FI-90014 Finland\\[\affilskip]
$^4$Instituto de Astronom\'ia, Universidad Nacional Aut\'onoma de M\'exico, Apdo. Postal 877, Ensenada, Baja California 22800, M\'exico
}
\begin{document}

\maketitle

\begin{abstract}
Rings and spiral arms are distinctive features of many galaxies, and their properties are closely related to the disk dynamics. 
They are often associated to stellar bars, but the details of this connection are far from clear. 
We study the pitch angles of spiral arms and the frequency and dimensions of inner and outer rings as a 
function of disk parameters and the amplitude of non-axisymmetries in the S$^4$G survey. 
The ring fraction increases with bar Fourier density amplitude: 
this can be interpreted as evidence for the role of bars in ring formation. 
The sizes of inner rings, normalised by the disk size, are positively correlated with bar strength: 
this can be linked to the radial displacement of the inner 4:1 ultra-harmonic 
resonance while the bar grows and the pattern speed decreases. 
The fraction of rings is larger in barred galaxies than in their non-barred counterparts, 
but still $\sim 1/3$ ($\sim 1/4$) of the galaxies hosting inner (outer) rings are not barred. 
The amplitudes of bars and spirals are correlated for all types of spirals. 
However, on average, the pitch angles of spiral arms are roughly the same for barred and non-barred galaxies: 
this questions the role of bars exciting spiral structure. 
We conclude that the present-day coupling of rings, spiral arms, and bars is not as robust as predicted by simulations.
%
\end{abstract}

\firstsection 

\section{Context and aims}

Do galactic bars drive the formation of rings and spiral arms? Simulations predict so 
\citep[e.g.][and references therein]{Sellwood,Athanassoula}:  
\begin{enumerate}
\item most rings form from interstellar gas collected near disk resonances \citep[e.g.][]{Schwarz81,Rautiainen}, under the action of bar gravity torques,
\item bars excite spiral density waves \citep[e.g.][]{Sanders}, 
\item rings/spirals can be made of orbits organised in tubes (invariant manifolds) originating from close to the 
bar ends \citep[e.g.][]{Romero}.
\end{enumerate}

Here, we address the coupling between bars, rings, and spirals in the 
\emph{Spitzer} Survey of Stellar Structure in Galaxies \citep[S$^4$G;][]{Sheth10}. 
We use 3.6 $\mu$m imaging for a parent sample of 1320 nearby galaxies with 
inclinations lower than $65^{\circ}$ \citep[according to][]{Salo15}, 
of which \citep[according to][]{Buta15} 825 are barred, 
465 host inner rings, 264 host outer rings, 
and 391 have measurements of the pitch angle (winding angle) of the spiral arms 
\citep[from][]{Herrera-Endoqui15,Diaz-Garcia19b}. 
\section{Strength of spiral arms and bars} 
\begin{figure}[h]
\begin{center}
\begin{tabular}{c c}
\includegraphics[width=0.55\textwidth]{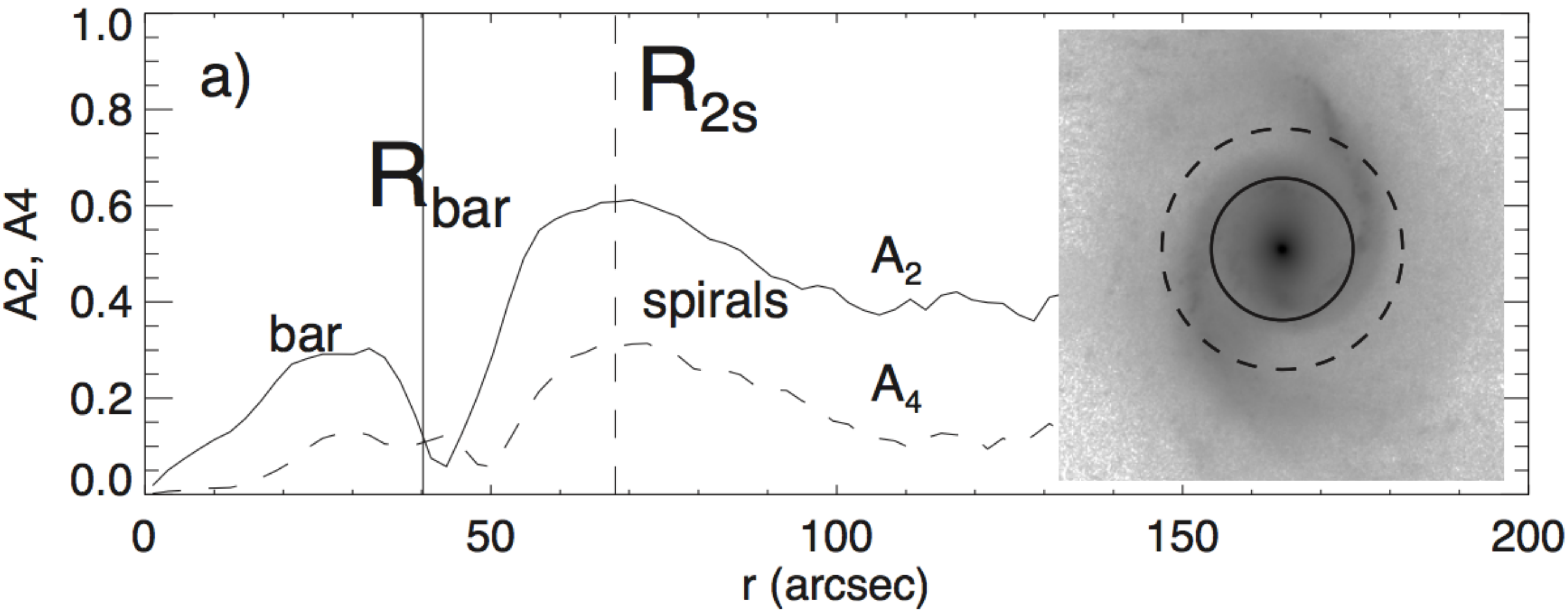}
\includegraphics[width=0.45\textwidth]{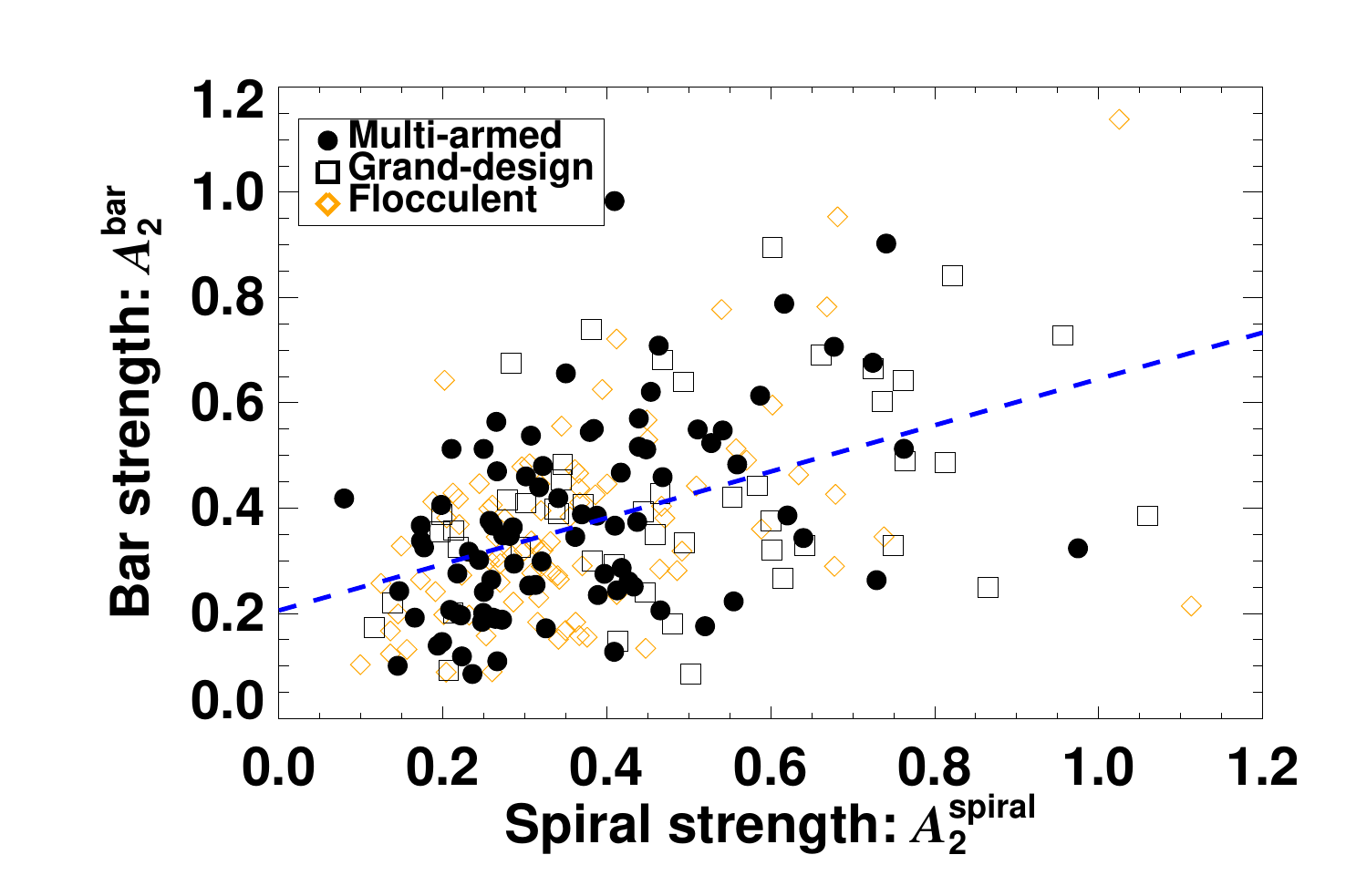}
\end{tabular}
\caption{
\emph{Left:} Example of bar/spiral force calculation for NGC$\,$1566 \citep[from][]{Salo10}. 
Radial profiles of the $m=2,4$ Fourier density amplitudes and de-projected $K_{S}$-band image (inner panel). 
The vertical lines indicate the bar length (solid) and the distance of maximum $A_{2}$ of the spirals (dashed) 
(these radii are also overlaid on the image). 
\emph{Right:} Spiral strength \citep[from][]{Diaz-Garcia19b} versus bar strength \citep[from][]{Diaz-Garcia16a}, 
measured from $A_2$ using 3.6 $\mu$m imaging. 
Different colours and symbols represent different types of spirals (see legend). 
The dashed blue line shows the linear fit to the cloud of points.
}
\label{Fig03}
\end{center}
\end{figure}
We calculate the amplitudes of non-axisymmetries from the maximum of the $m=2$ normalised Fourier density amplitudes ($A_{2}$) 
associated to the bar \citep[$A_{2}^{\rm bar}$, from][]{Diaz-Garcia16a} 
and to the spiral arms \citep[$A_{2}^{\rm spiral}$, from][]{Diaz-Garcia19b} (left panel on Fig.~\ref{Fig03}), 
using $3.6\,\mu$m S$^4$G images.

We confirm that the strengths of bars and spirals are correlated (right panel of Fig.~\ref{Fig03}) \citep[e.g.][]{Salo10}. 
This either i) supports the role of bars driving the formation of spirals \citep{Sanders} 
or ii) indicates that the disks that are prone to the formation of strong bars are also more reactive 
to the development of spirals of large amplitudes \citep{Salo10,Diaz-Garcia16b}, while the correlation does 
not necessarily imply causation. The latter interpretation (ii) is favoured by the observed coupling of 
the amplitudes of bar and arms even in flocculent spirals 
or when only the outermost segments of multi-armed galaxies are analysed 
(most likely, bars are not responsible for exciting flocculent arms or outer spiral modes) 
\citep{Diaz-Garcia19b}. 
\begin{figure}[h]
\begin{center}
\begin{tabular}{c c}
\includegraphics[width=0.55\textwidth]{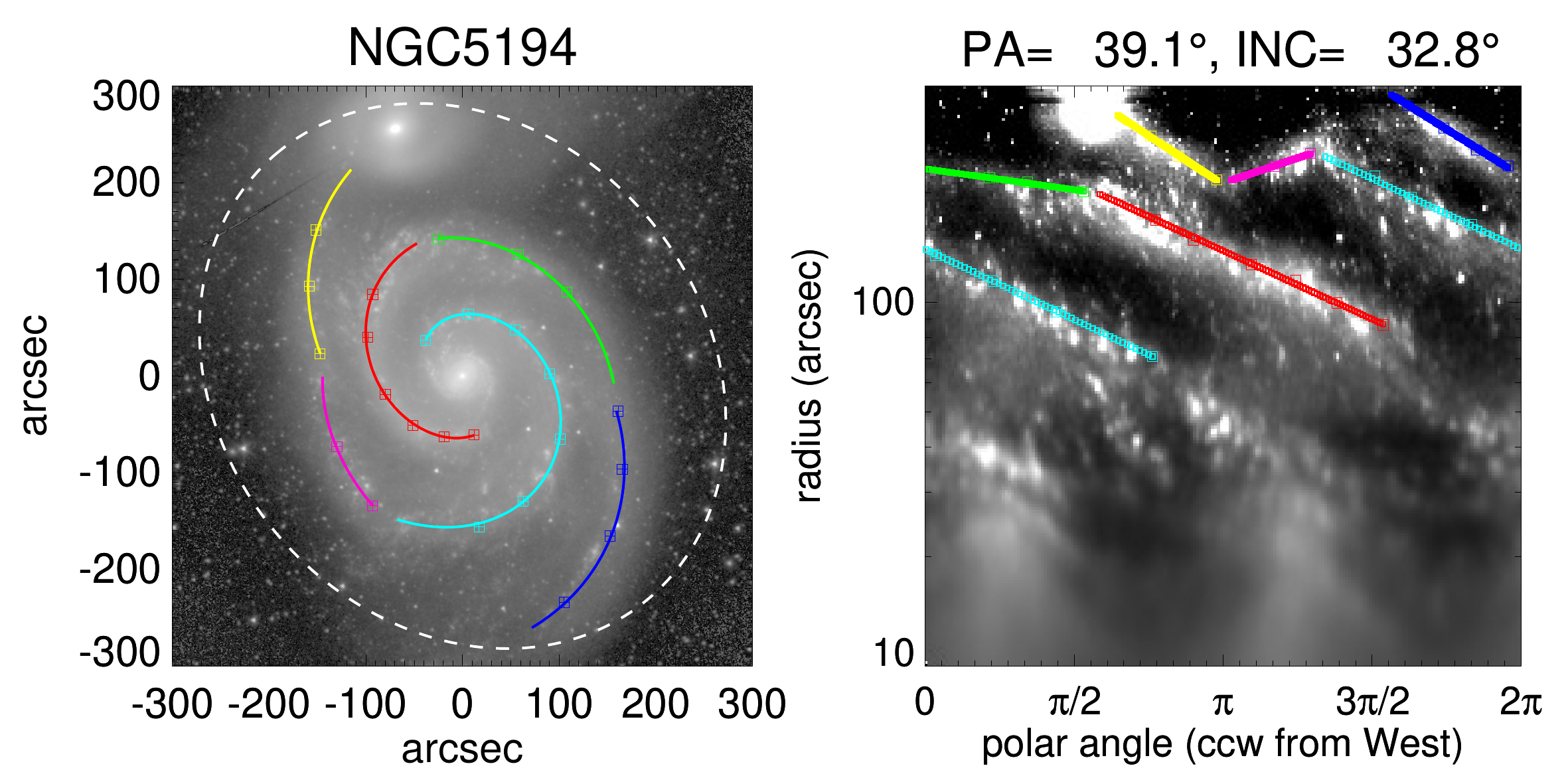}
\includegraphics[width=0.45\textwidth]{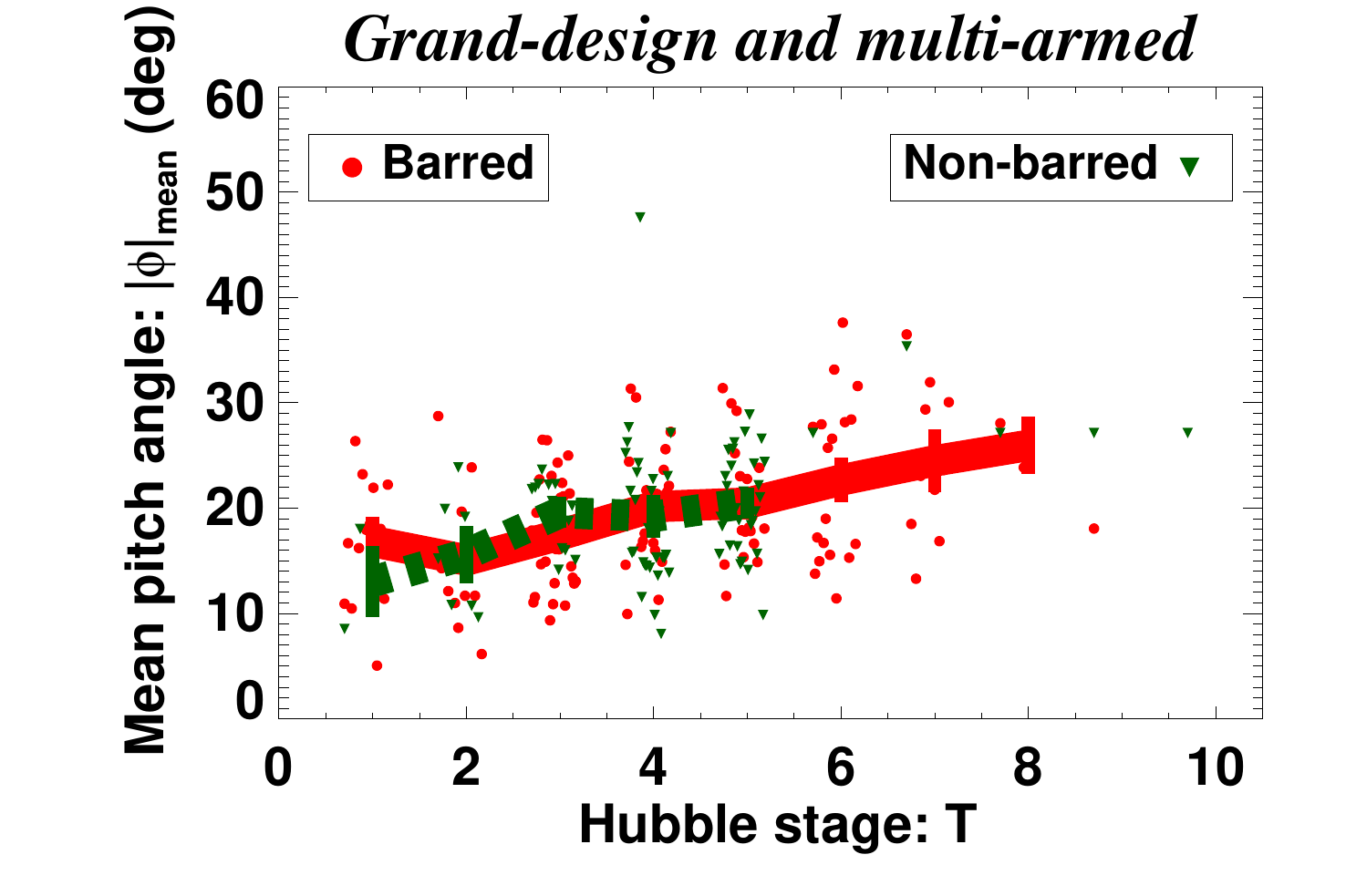}
\end{tabular}
\caption{
The 3.6 $\mu$m image of M$\,$51 in the sky plane (\emph{left panel}) 
and the logarithmic polar plot de-projected to the disk plane (\emph{central panel}). 
Overlaid with different colours are the fitted logarithmic spiral segments by \cite{Herrera-Endoqui15}. 
In the \emph{right panel} we show the mean pitch angle of the galaxy 
as a function of the integer value of the revised numerical Hubble stage, 
for all the grand-design and multi-armed spirals in our sample. 
The running mean and standard deviation of the mean are shown for the barred (red) and non-barred (green) galaxies.
}
\label{Fig02}
\end{center}
\end{figure}
\section{Pitch angle of spiral arms in barred and non-barred galaxies} 
\cite{Herrera-Endoqui15} calculated the pitch angles of spiral arm segments using 3.6~$\mu$m S$^4$G photometry. 
They visually marked points tracing the spiral segments and performed a linear fit in the disc plane in a polar coordinates, 
where logarithmic arms appear as straight lines (left and central panels of Fig.~\ref{Fig02}). 
In order to parameterise the global winding of the spirals, 
we calculate the mean of the absolute value of the pitch angle measurements of logarithmic segments 
($|\phi|_{\rm mean}$)\citep{Diaz-Garcia19b}. 

For grand-design and multi-armed spirals, 
the global pitch angle increases with increasing Hubble type ($T$) (right panel of Fig.~\ref{Fig02}), as expected, 
but with a large scatter. 
Interestingly, the distribution of pitch angles for barred and non-barred galaxies is roughly the same when 
$1 \le T\le 5$: this questions the role of bars driving spiral density waves.
\section{Interplay between bars and rings}
\underline{Rings are more common in galaxies with stronger bars}:
The fraction of inner and outer rings \citep[as identified by][]{Buta15} increases with 
increasing bar Fourier density amplitude (upper panels of Fig.~\ref{Fig12}): 
this can be interpreted as evidence for the role of bars in ring formation \citep[][]{Diaz-Garcia19a}. 
However, $\sim 1/3$ ($\sim 1/4$) of the galaxies hosting inner (outer) rings are not barred 
(lower panels of Fig.~\ref{Fig12}), and thus i) some bars dissolve after ring formation (implausible based on simulations), or 
ii) other mechanisms may be responsible for ring creation (e.g. spirals or interactions).
\underline{Concurrent growth of inner rings and bars}: 
\begin{figure}[b]
\begin{center}
\begin{tabular}{c c c c}
 \includegraphics[width=0.41\textwidth]{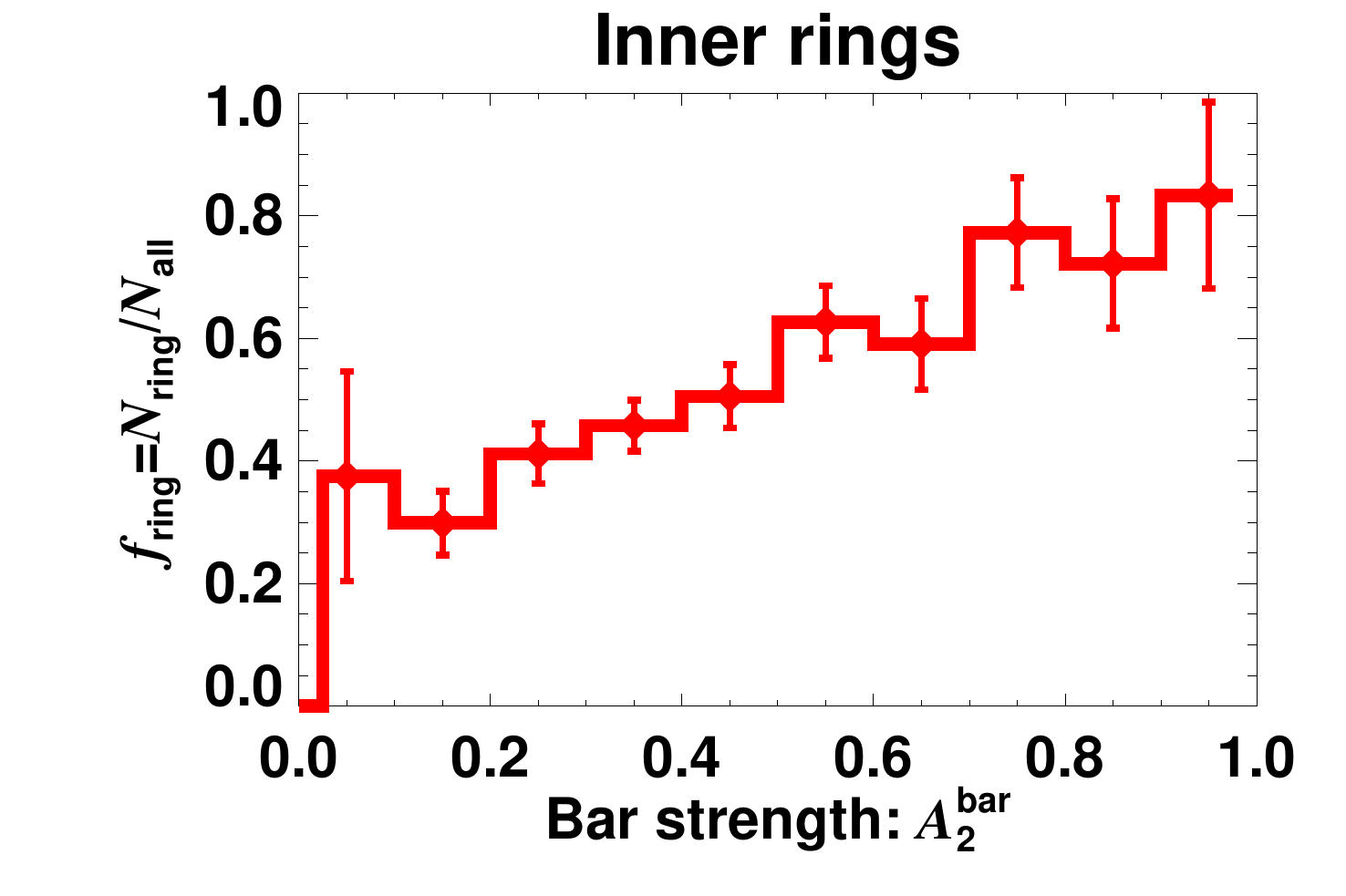}
 \includegraphics[width=0.41\textwidth]{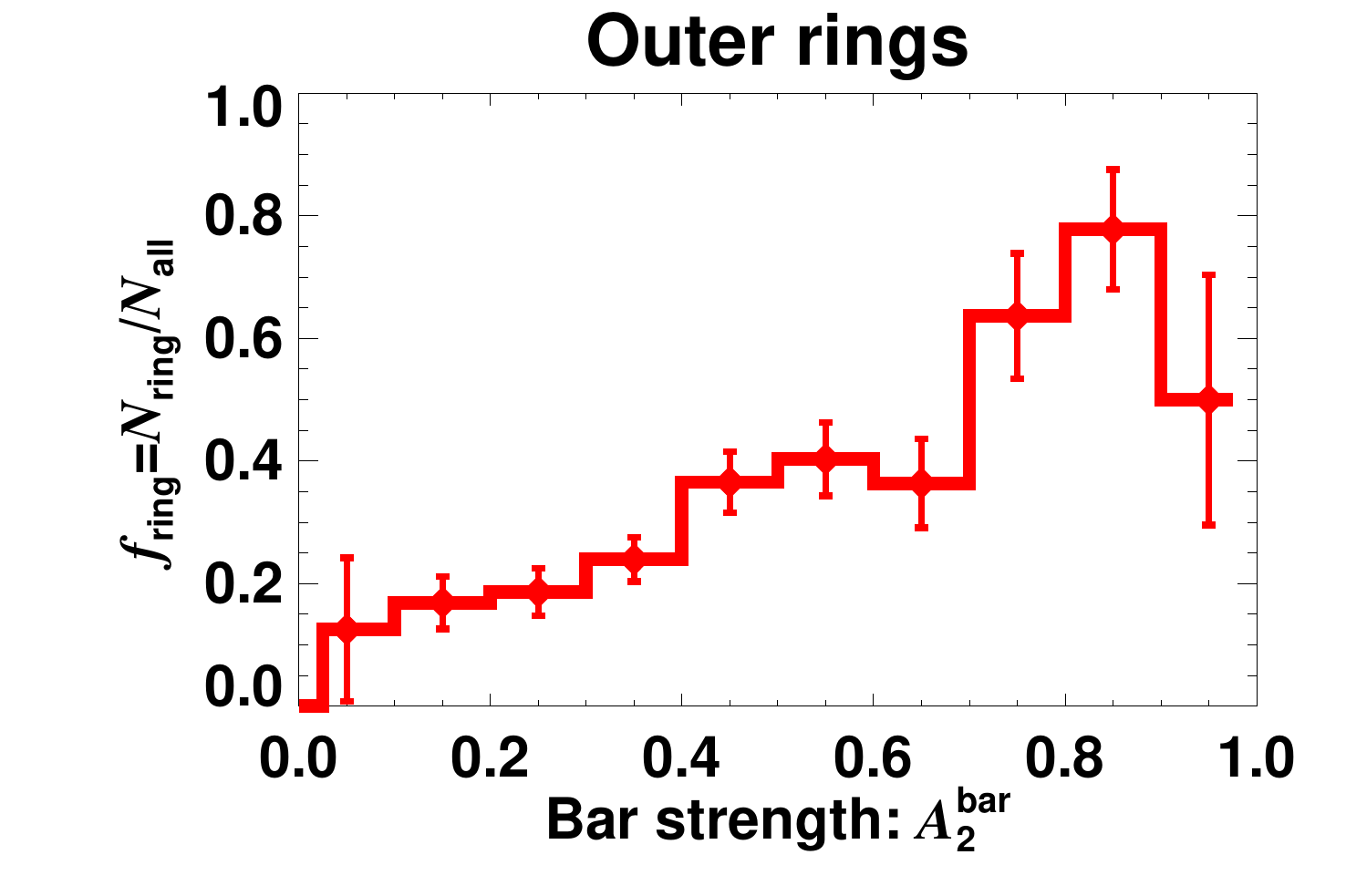}\\
  \includegraphics[width=0.41\textwidth]{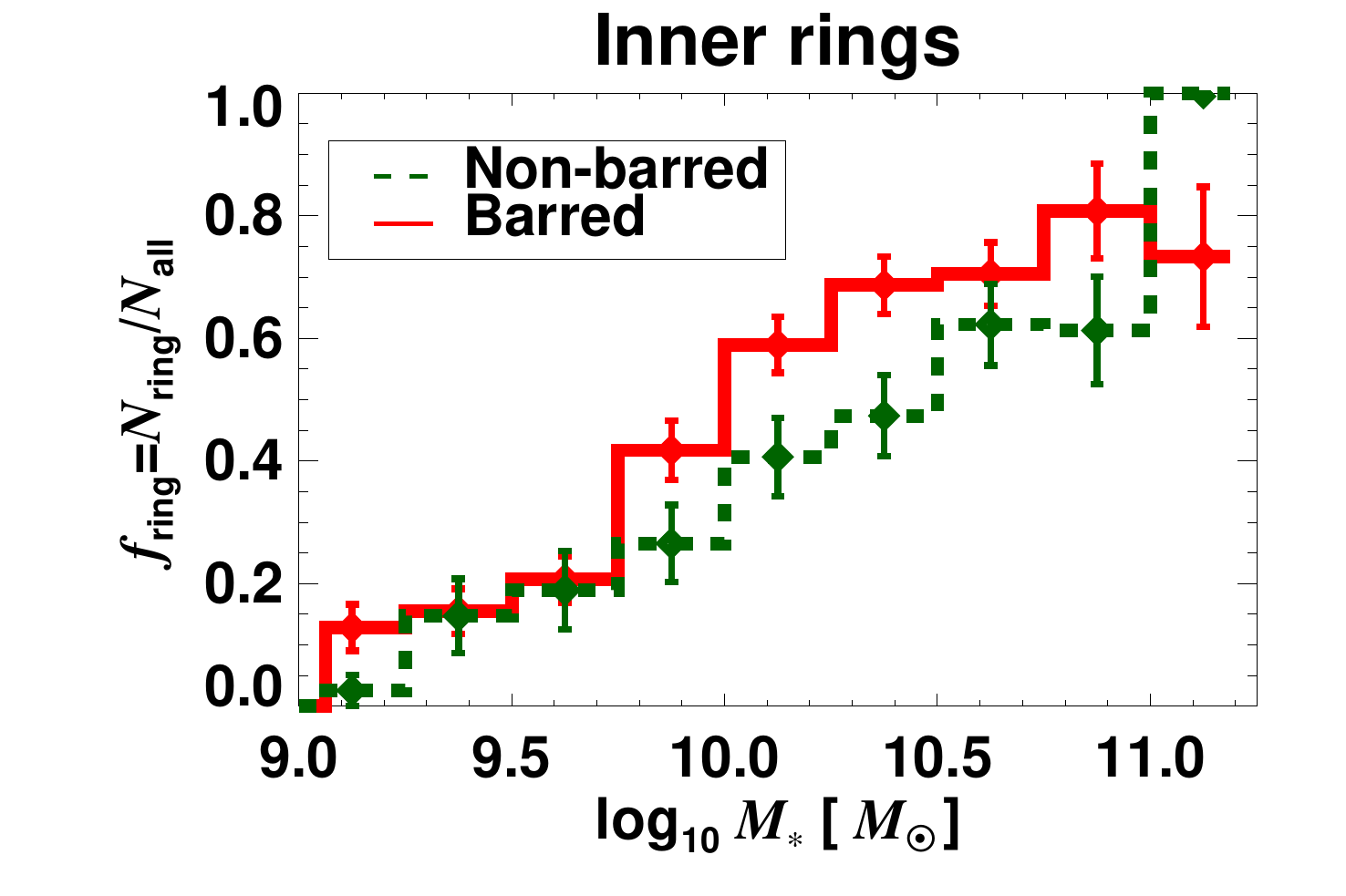}
 \includegraphics[width=0.41\textwidth]{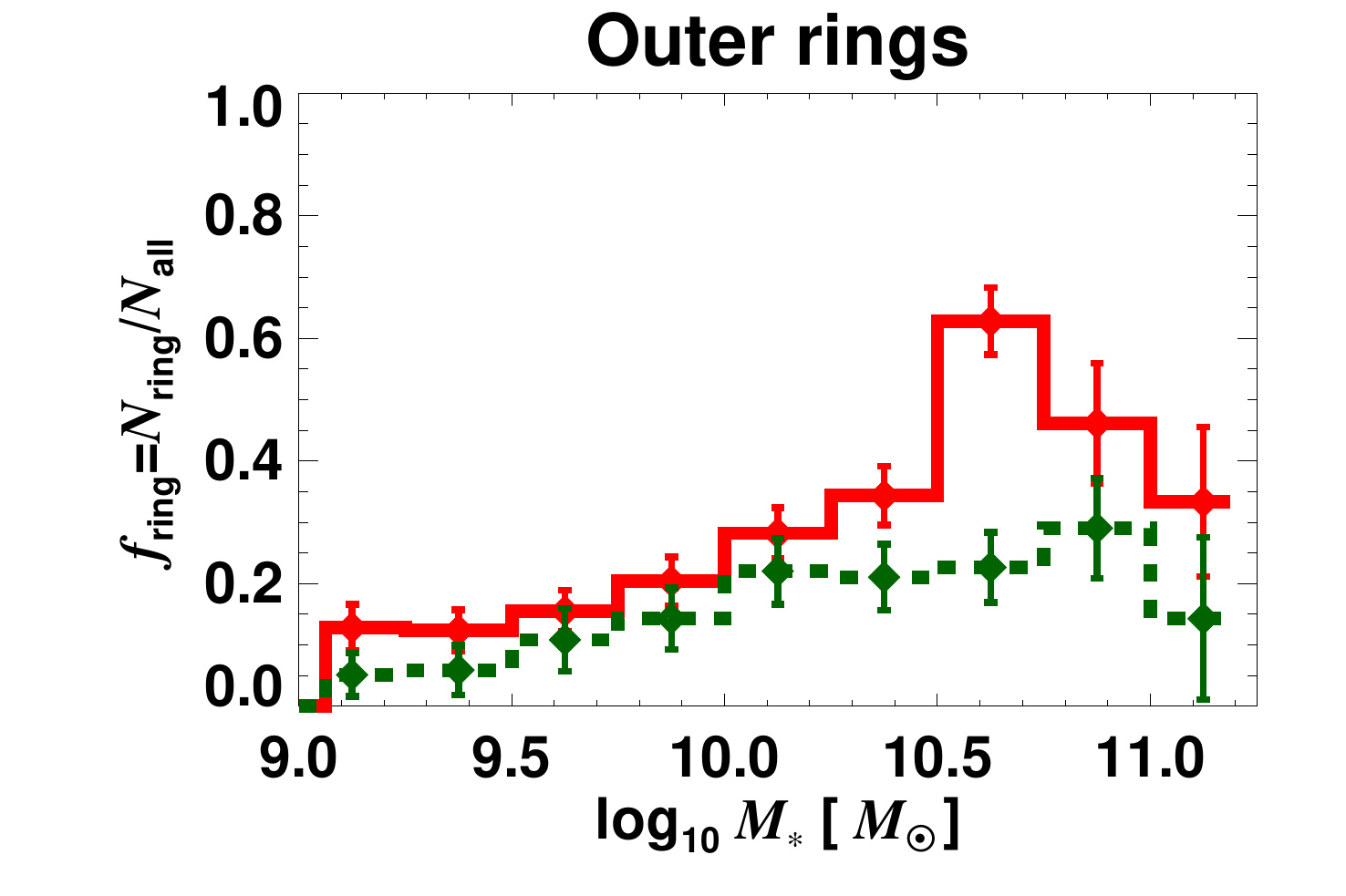}
\end{tabular}
\caption{
Fraction of inner (\emph{left}) and outer (\emph{right}) rings as a function of 
$m=2$ bar Fourier amplitude (\emph{upper panels}) and total stellar mass of the galaxies 
(\emph{lower panels}, including non-barred galaxies). 
Error bars correspond to binomial counting errors. 
The fraction of rings increases with increasing $M_{\ast}$ and $A_{2}^{\rm bar}$. 
The fraction of inner (outer) rings in barred galaxies is $1.41 \pm 0.12$ ($1.88 \pm 0.25$) 
times larger than in their non-barred counterparts \citep[see also][]{Comeron14}.
}
  \label{Fig12}
\end{center}
\end{figure}
\begin{figure}[h]
\begin{center}
\begin{tabular}{c c}
  \includegraphics[width=0.55\textwidth]{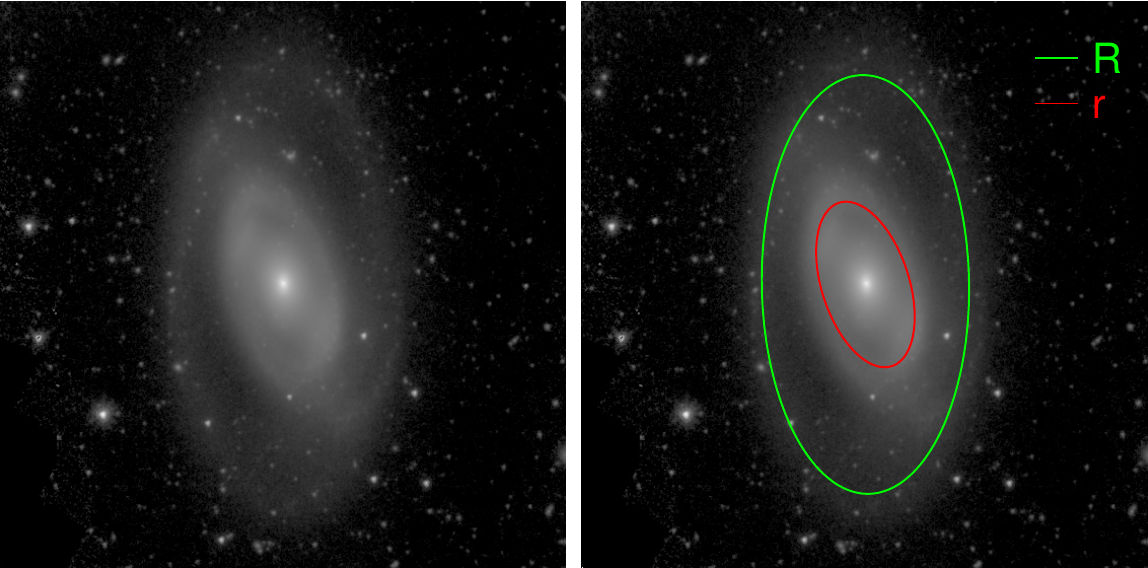}
  \includegraphics[width=0.45\textwidth]{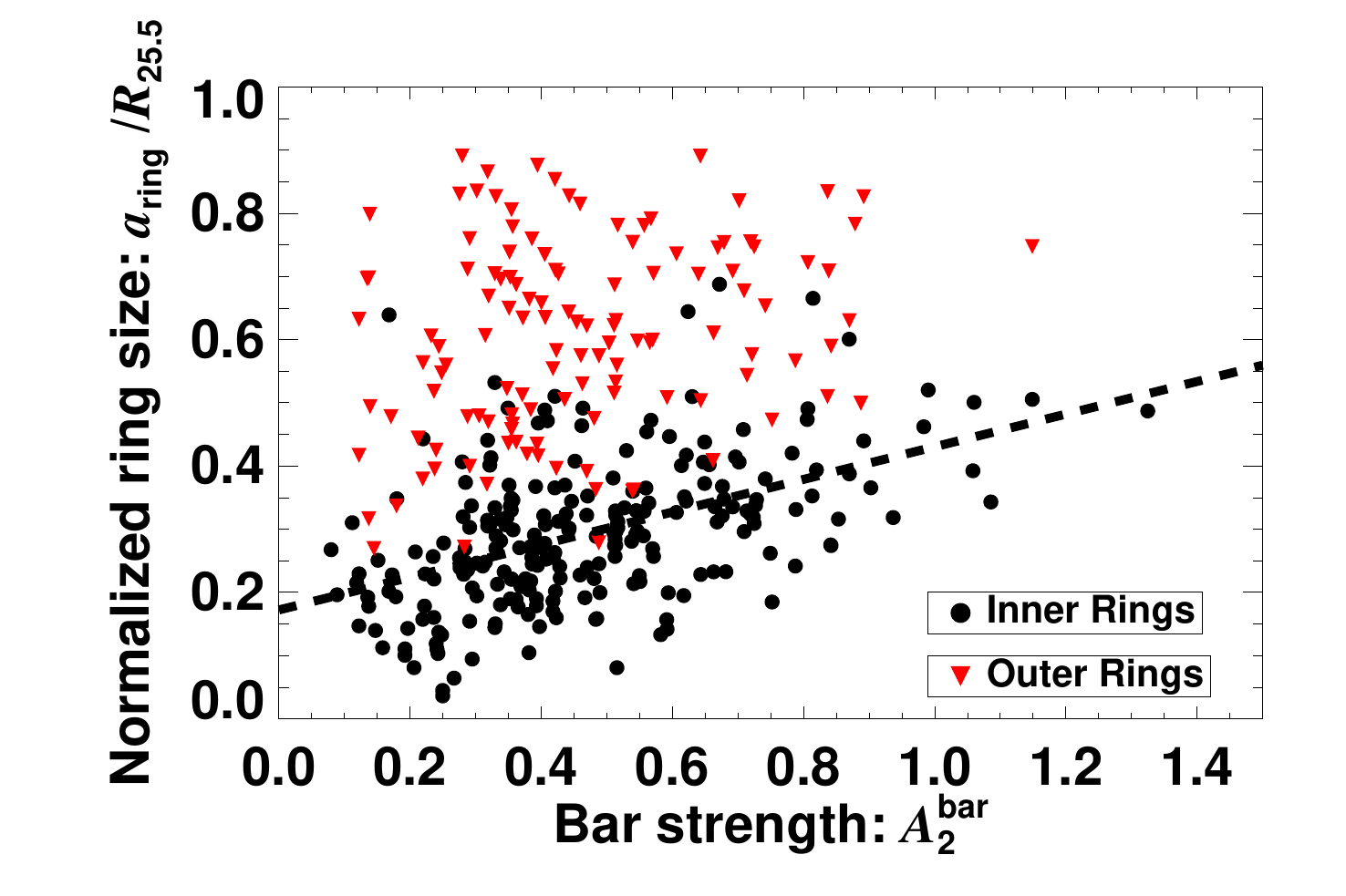}
\end{tabular}
\caption{
\emph{Left and central panels:} 
Measurements of the sizes and axial ratios of the inner (red) and outer (green) rings hosted by NGC$\,$1350 
\citep[from][]{Herrera-Endoqui15}. The outline of the ridge of the rings was visually marked 
and fitted with an ellipse on the 3.6 $\mu$m images. 
\emph{Right panel:} 
For a sample of non-highly inclined disk galaxies, de-projected semi-major axis of rings, normalised to $R_{25.5}$ 
\citep[isophotal radii at 25.5~mag~arcsec$^{-2}$ at $3.6\,\mu$m, from][]{Mateos}, 
versus the $m=2$ bar Fourier amplitude \citep[][]{Diaz-Garcia19a}, separating inner and outer rings (see legend). 
The dashed line shows the linear fit to the data cloud for inner rings.
}
\label{Fig01}
\end{center}
\end{figure}
We use measurements of sizes and axial ratios of inner and outer rings (left and central panels of Fig.~\ref{Fig01}) 
from \cite{Herrera-Endoqui15}. The sizes of inner rings are correlated with bar strength (right panel of Fig.~\ref{Fig01}): 
we interpret this as a consequence of the concurrent growth of bars (whose strength and length increase in time) 
and inner rings, as the inner 4:1 ultraharmonic resonance moves outwards in the disc while the pattern speed decreases 
\citep[][]{Diaz-Garcia19a}. Outer ring sizes do not correlate with bar strength: 
this is probably linked to the larger timescales required for outer ring formation from gas redistributed by bars, 
whose potential might have changed in time.
\section{Conclusions}
\begin{itemize}

\item Bars play a role in inner/outer ring formation. 
However, the coupling between bars and rings is not as strong as expected from numerical models \citep[][]{Diaz-Garcia19a}.

\item Disks that are prone to the development of strong bars are also reactive to 
the formation of prominent spirals. This does not imply that spirals are bar-driven. 
Statistically, barred and non-barred galaxies have similar pitch angles \citep[][]{Diaz-Garcia19b}.

\end{itemize}
\end{document}